\documentstyle[11pt,newpasp,twoside,epsf]{article}

\begin{document}

\title{The Central kpc of Galaxy Bulges}

\author{Marc Balcells}

\affil{Instituto de Astrof\'\i sica de Canarias, 38200 La Laguna, 
Tenerife,
Spain}

\begin{abstract}
We study the innermost regions of bulges with surface
brightness data derived from combined HST/NICMOS and ground-based NIR
profiles.  Bulge profiles to 1-2 kpc may be fit with Sersic laws, and
show a trend with bulge-to-disk ratio:  low-B/D bulges are roughly
exponential, whereas higher-B/D bulges show increasing Sersic shape 
index $n$,
indicating higher peak central densities and more extended brightness
tails.  $N$-body models of accretion of satellites onto disk-bulge-halo
galaxies show that satellite accretion contributes to the
increase of the shape index $n$ as the bulge grows by accretion.  
The $N$-body
results demonstrate that exponential profiles are fragile to merging,
hence bulges with exponential surface brightness profiles cannot have
experienced significant growth by accretion of dense satellites.   
\end{abstract}

\section{Introduction}

Bulges are the dominant mass and light component of the central kpc in
early- and intermediate-type disk galaxies, thus it is appropriate to
address their formation and evolution at this conference.  Bulges are
often assimilated to small ellipticals residing in the centers of
spirals, and indeed the similarities include structure, colors,
populations (Wyse, Gilmore, \& Franx 1997), age (Peletier et al.\  
1999) and position in the Fundamental Plane (Falcon-Barroso, Peletier,
\& Balcells 2001).  However, the intimate scaling of disk and bulge
luminosity and structural sizes along the Hubble sequence (de Jong
1996, Courteau \& de Jong 1997, Graham 2001) and the similarity of
their colors (Peletier \& Balcells 1997) indicate that bulge and disk
have known about each other during their formation and growth, and
that understanding a galaxy amounts to more than understanding its
components.

An increased attention to bulges during the nineties has allowed us 
to improve on the picture that bulges are described by the $r^{1/4}$ 
surface brightness law, that they are isotropic oblate rotators 
(Kormendy \& Illingworth 1983) and that their populations are of high 
metallicity (Rich 1988).  As first noted for the Milky Way bulge by 
Kent, Dame, \& Fazio (1991), small bulges have exponential surface 
brightness profiles (Andredakis \& Sanders 1992, Andredakis, 
Peletier, \& Balcells 1995, de Jong 1996).  Box- and peanut-shaped 
bulges have cylindrical rotation (Shaw, Wilkinson, \& 
Carter 1993) and show the kinematic signature of barred dynamics 
(Kuijken \& Merrifield 1995; Bureau \& Freeman 1999).  The stellar 
populations of bulges have below-Solar metallicities (Balcells \& 
Peletier 1994; Sadler, Rich, \& Terndrup 
1996).  Recently, a strong correlation has been found between the 
mass of central black holes and the velocity dispersion of the bulges 
they reside in (Magorrian et al.\ 1998; Merrit \& Ferrarese 2001).  

On the theoretical side, $N$-body models have shown that vertical 
buckling instabilities in bars may form a peanut-shaped bulge-type 
object at the center of disk galaxies (Combes \& Sanders 1982; 
Hassan, Pfenniger \& Norman 1993).  Bulges 
may therefore be by-products of disk secular evolution.  That bulges 
do not always precede disk formation is evidenced by the fact that 
bulge-less galaxies do exist.  

Their high central densities suggests that bulge formation involves 
violent, efficient starbursts, while their z-extent suggests violent 
dynamical processes involving disk instabilities or merging.  

These findings have added new ingredients to the basic questions on 
bulges:  how old are bulges; do bulges form 
before/after/simultaneously with their disks; do bulges grow and does 
the galaxy evolve along the Hubble sequence; is there a dichotomy, 
such that old bulges are old collapse/merger products and small 
bulges are young products of disk instabilities; is bulge formation 
and growth connected to black hole formation and growth.  

To some degree any study of bulges addresses what we understand by 
bulges.  The term 'bulge' refers to the swelling at the center of the 
spiral disk, hence to the extended structure.  But because this 
swelling can only be observed in the most edge-on cases, in practice 
bulges are taken to be the central brightness increase above the 
inward extrapolation of the exponential disk profile.  Because the 
disk may contain structural components embedded in the bulge, eg. 
bars, these two definitions need not describe the same structures.  

In this paper I review the systematics of bulge surface brightness 
profiles, as derived on a sample of early- to intermediate-type 
inclined galaxies, and describe the results of $N$-body merger studies 
of the effects of satellite accretion on the surface brightness 
profiles.  I then present our work on combined HST+ground based NIR surface 
brightness profiles, which map bulges from $\sim$20 pc to a few kpc.  

\begin{figure}
\plotfiddle{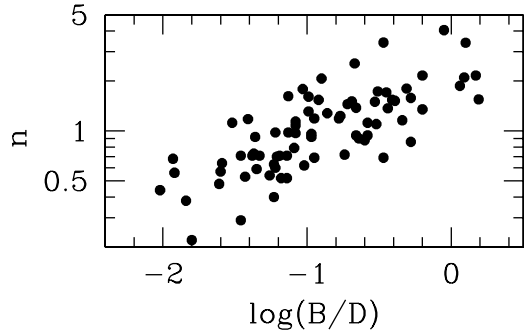}{5cm}{0}{50}{50}{-180}{-170}
\caption{The $K$-band correlation of shape index $n$ vs. bulge-to-disk ratio 
for 86 bulges from de Jong (1996) fitted by Graham (2001). }
\label{MB1:Fig:NvsBD}
\end{figure}

\section{The Sample}
We work on a diameter limited, statistically complete sample of 30 
unbarred, inclined ($i > 50\deg$) galaxies from the UGC (Nilson 1973), 
of types from S0 to Sbc.  Published papers on this sample address optical colors, NIR surface brightness profiles, bulge-disk decomposition, disk-bulge colors, ages, and Fundamental Plane (Balcells \& Peletier 1994; Peletier \& Balcells 1996; Peletier \& Balcells 1997; Peletier et al.\ 1999; Falc\'on-Barroso, Peletier, \& Balcells 2001; Balcells et al. 2001). 
All of our ground-based imaging data is made freely available in an electronic atlas (Peletier \& Balcells 1997). 

\section{Dichotomy vs continuum of shapes}
The departure of bulges from the $r^{1/4}$ surface brightness law has 
been approached in different ways by different authors.  The most 
convenient method is to use Sersic's (1968) law, 

\[ I(r)=I_{e} \cdot
\exp{\left\{-b_{n}\cdot\left[\left(\frac{r}{r_{e}}\right)^{1/n}-1\right]\right\}},\]

\noindent with $b_{n}\approx 1.9992\cdot n-0.3271$ (Caon, Capaccioli, \& d'Onofrio 
1993).  We refer to $n$ as the shape index or concentration 
index. Exponential profiles are 
obtained setting $n=1$, and $n=4$ gives the $r^{1/4}$ law.   One 
approach is to choose the best fitting $n$ from the set $n=1,2,4$ (de 
Jong 1996; Graham \& Prieto 1999).  A more restricted approach is to 
simply classify objects into two groups, the "exponential bulges" 
($n=1$) and the "$r^{1/4}$ bulges" ($n=4$; Carollo 1999).  Such 
approaches may be over restrictive as, in fact, $n$ seems to be 
continuously distributed (Andredakis et al.\ 1995; Graham 2001).  The 
latter two papers demonstrate that $n$ strongly correlates with B/D 
and with the total luminosity of the bulge, in the sense that 
late-type galaxies are constrained to $n\la$ 1 while early-type 
spirals show $n$ reaching up to $n \sim 5$ (Fig.~\ref{MB1:Fig:NvsBD}).  
This result holds for both blue and NIR passbands and is irrespective 
of whether the bulge-disk decomposition is done on the 1D surface 
brightness profile (Graham 2001) or in a model-independent fashion 
using ellipticity signatures of the bulge (Andredakis et al.\ 1995).  
Graham (2001) shows that choosing the wrong $n$ leads to distortions 
on other parameters such as the disk effective radius and the B/D.  

The trend of $n$ vs luminosity in bulges follows a similar trend 
found by various authors for ellipticals (eg. Caon et al.\ 1993, 
Trujillo et al.\ 2001) and spans a range of 8 magnitudes. 

What clues does that trend contain about the formation of ellipticals 
and bulges? Andredakis (1998) studied the effects of the disk 
potential on the surface brightness profile of the bulge, using 
$N$-body simulations of the adiabatic growth of a disk potential on a 
pre-existing $r^{1/4}$ bulge.  He shows that the disk potential 
drives the $r^{1/4}$ profile toward lower $n$.  However, the effect 
saturates around $n=2$, demonstrating that the disk potential alone 
cannot turn $r^{1/4}$ spheroids into exponentials.  

\section{Growth of bulges: testing the accretion hypothesis}

The continuity of the trend along the bulges and ellipticals suggests 
a process that operates on all spheroids independent of the 
presence of a massive disk. Aguerri, Balcells \& Peletier (2001) have 
used $N$-body merger simulations to investigate the effects of the 
accretion of satellites on the surface brightness profiles of 
bulges.  They study the scenario in which initially small, 
exponential bulges grow by accretion of 
satellites.  The aim is to see whether bulge growth is accompanied by 
an increase of the shape index $n$.  This scenario fits in with the 
fact that low-luminosity bulges are those with exponential profiles.  

\begin{figure}
\plotfiddle{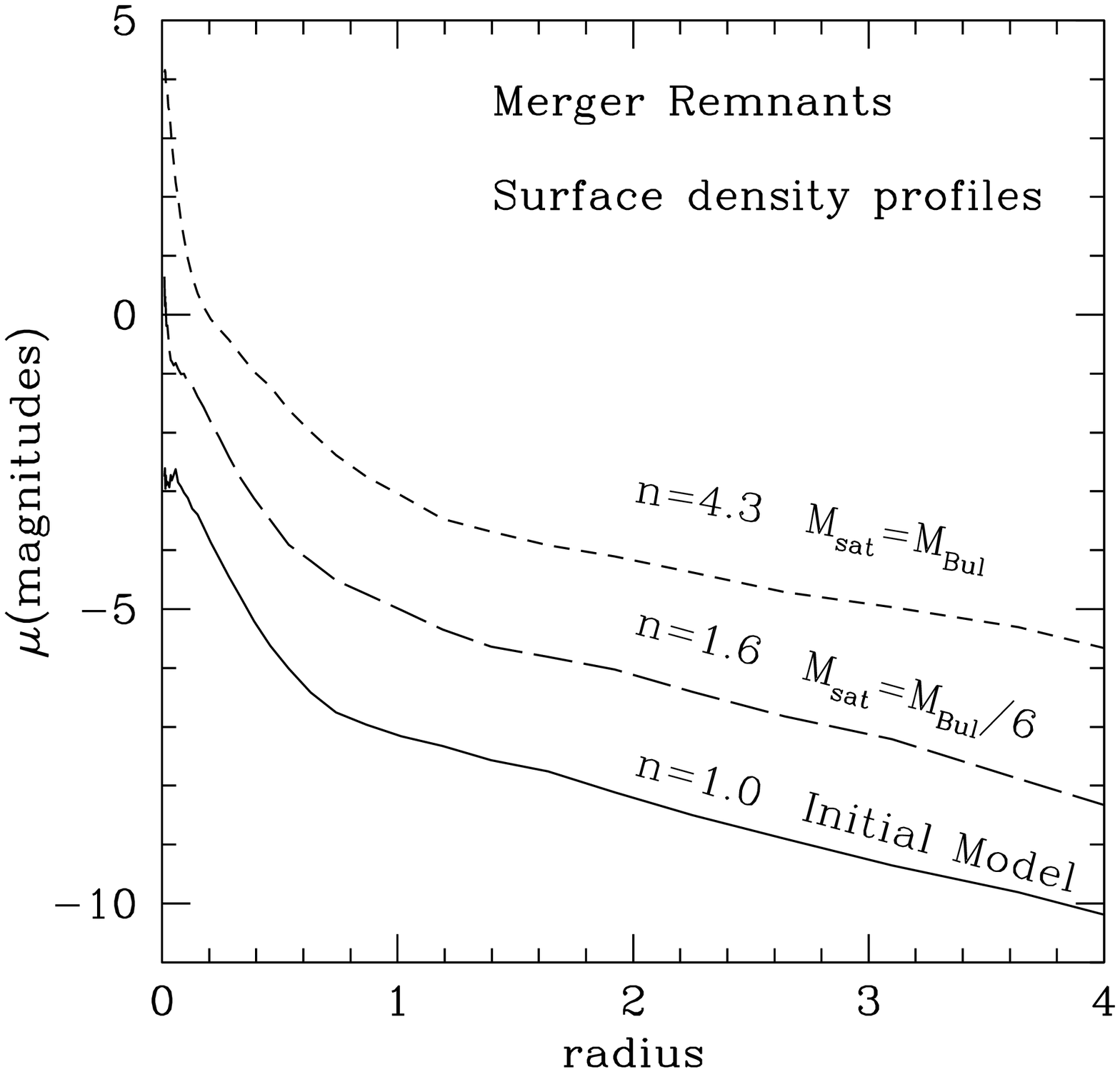}{4cm}{0}{25}{25}{-190}{-75}
\plotfiddle{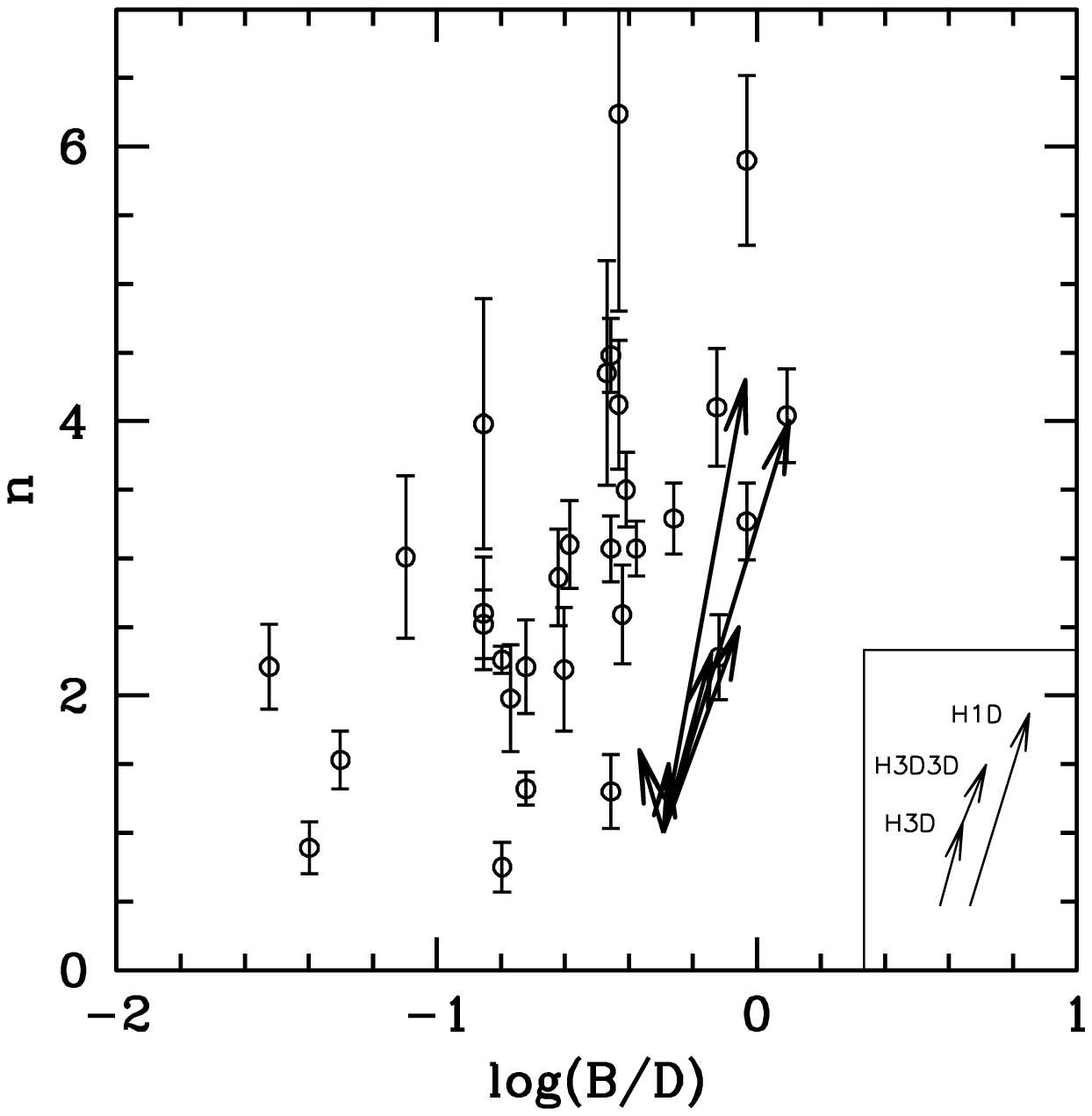}{0cm}{0}{37}{37}{-40}{-100}
\caption{{\it (a)} Sample surface density profiles of the luminous matter in $N$-body merger remnants, and in the initial spiral galaxy model. 
{\it (b)} Growth vectors in the $n$ vs B/D plane, and the data from Andredakis et al.\ (1995). {\it Inset: }  The cumulative effect of a second merger.  }
\label{MB1:Fig:Simul}
\end{figure}

Simulations use the Kuijken \& Dubinski (1995) disk-bulge-halo galaxy
model as the primary, and either a King or a Hernquist (1990) sphere
for the satellite.  Evolution is computed with Heller's treecode
(Heller \& Schlosman 1994), using N=110,000 particles.  Simulation
details are given in the reference above.  Experiments explore a range
of satellite masses, prograde and retrograde merger orbits and two
different satellite models.  Secondary mergers, as well as one
low-density satellite case, are also studied.  In all cases, the
system is allowed to merge and relax, after which the face-on surface
density profile is extracted and fitted with exponential plus Sersic
laws.  The fit procedure is similar to that done on the profiles of
real galaxies, and produces best-fit parameters $R_e$, B/D, disk
scale length $h$ and central surface brightness $\mu_0$, in addition 
to the bulge shape index $n$.  This allows one to draw growth vectors in the 
$n$-B/D plane.  Examples of merger remnant surface density profiles are shown in Figure~\ref{MB1:Fig:Simul}a.

Whenever an accreted satellite reaches the bulge, $n$ rapidly increases as B/D grows.  Figure~\ref{MB1:Fig:Simul}b shows growth vectors for bulges in the
merger simulations, together with the distribution of bulges in
Andredakis et al.\  (1995).  Roughly a 25\%\ in B/D is
associated to an increase $\Delta(n)=1$, and a satellite as massive as
the bulge results in an $r^{1/4}$ final bulge.  To first order the
effect is cumulative, ie a second accretion causes the same increment
in $\log(B/D)$ and $n$ as the first (see inset to
Fig.~\ref{MB1:Fig:Simul}b).  The evolution drawn by the mergers in
broad terms reproduces the distribution of observed bulges, although 
it is steeper in the simulations.  This may be due in part 
to the fact that both the initial bulges and the satellites are more 
massive in the simulations than in many of the observed galaxies, as 
well as to the dense nature of the satellites used in this set of 
simulations.  

The increase of $n$ in the models is due to both mass deposition and
to heating of the bulge mass distribution.  High-$n$ Sersic profiles
have both high central peaks and extended tails;  in the models, it is the
tails which drive the increase of $n$, as the models lack
central resolution.  We note that mass pile-up at the center includes 
disk material drawn in by transient bar distortions during the 
merger.  The standard bulge-disk decomposition, which fits the disk 
with a simple exponential, assigns some of this central disk material to the 
bulge.  This may contribute to the color similarity between the bulge 
and the inner disk (Peletier \& Balcells 1996).  

The bulge only evolves if and when the satellite reaches the bulge before 
disrupting.  Low density satellites deposit the 
mass in the halo and the disk, and the bulge remains unaffected.  It 
will be useful to investigate satellites with a range of density 
profiles, which may show behaviours intermediate between the 
high-density cases which fully merge with the bulge, and the 
low-density cases which leave the bulge unaffected.  

As expected, the accretion of the dense satellite heats up the disk
(eg.  Quinn, Hernquist, \& Fullagar 1993).  Aguerri et al.\ (2001)
show that the remnant disk presents a significant increase in scale
length $h_D$ in addition to an increase of scale height.  Such
disks resemble thick disks observed in early-type disk galaxies (eg. 
de Grijs \& Peletier 1997), hence satellite accretion may explain both the high-$n$ bulges and the thick disks of early-type spirals.  
A match to real galaxies presupposes the formation of a new thin disk out of the gas left over in the merger (Kauffmann, Guiderdoni, \& White 1994), implying that our scenario is one of bulge-before-the-disk, despite the existence of a disk before the merger. 

Further simulations, with higher central resolution and with a range 
of satellite densities, should provide additional tests on this 
hypothesis.  

\section{Bulge structure down to 20 pc}

Elsewhere in this volume, Balcells et al.\ (2001) present 
NIR ($\sim H$)-band surface brightness profiles obtained by matching 
HST/NICMOS F160W profiles to ground-based $K$-band data, for a fraction
of the Balcells \& Peletier (1994) sample.  Adding the interesting 20 pc -- 200
pc radial range, the profile fits give systematic higher $n$, but still show a dependency of $n$ with B/D.  The profiles in the inner arcsec appear as clean power laws reminiscent of those found by Lauer et al.\ (1995) for ellipticals.

\acknowledgements The contents of this paper draws in part from collaborative work with Reynier Peletier and Alfonso L. Aguerri.  I acknowledge stimulating discussions on bulges with A. Graham, I. Trujillo, P. Erwin and L. Sparke.  Thanks are due to the organizers of this interesting conference.  This research has made use of the ADS and the LEDA extragalactic database.

\end{document}